\documentclass[a4paper,12pt]{elsarticle}

\usepackage[top=0.5in,bottom=0.75in,left=0.6in,right=0.6in]{geometry}
\usepackage{graphicx}
\usepackage[colorlinks=true,linkcolor=blue,citecolor=blue,urlcolor=blue]{hyperref}
\usepackage[figurename=Fig.]{caption}
\usepackage[tablename=Tab.]{caption}
\usepackage{color}
\usepackage{float}
\usepackage{amssymb}
\usepackage{amsmath}
\usepackage{mathrsfs}
\usepackage{ textcomp }

\usepackage{cases}
\usepackage{upgreek}
\usepackage{makecell}
\usepackage{pdflscape}
\usepackage{subfigure}
\usepackage{pdfpages}
\usepackage{lineno}
\usepackage{makeidx}
\usepackage{bm}
\usepackage{xstring}
\modulolinenumbers[1]
\usepackage{epstopdf}
\usepackage{lipsum}
\usepackage{slashed}
\usepackage{multicol}
\usepackage{setspace}
\usepackage{booktabs}
\usepackage[table]{xcolor}
\usepackage{multirow}

\usepackage[separate-uncertainty]{siunitx}

\newcommand{\unseen}[1]{\color{white}{#1}}
\newcommand{\sims}{\,\raisebox{0.5ex}{\texttildelow}\,}
\newcommand{\eref}[1]{(\ref{#1})}
\newcommand{\w}{\omega}
\newcommand{\vo}{\vec{o}\@ifnextchar{^}{\,}{}}
\def\up{\mathrm}
\def\degree{^\circ}

\newcommand{\doiref}[2]{\href{http://dx.doi.org/#1}{#2}}

\graphicspath{{PDF/}}
\definecolor{myblue}{RGB}{65,111,166}
\definecolor{myred}{RGB}{168,66,63}
\definecolor{mygreen}{RGB}{134,164,74}
\definecolor{mypurple}{RGB}{110,84,141}
\definecolor{myindigo}{RGB}{61,150,174}
\definecolor{myorange}{RGB}{218,129,55}
\definecolor{mylightblue}{RGB}{142,165,203}

\makeatletter
\def\ps@pprintTitle{%
  \let\@oddhead\@empty
  \let\@evenhead\@empty
  \let\@oddfoot\@empty
  \let\@evenfoot\@oddfoot
}
\makeatother



\biboptions{numbers,sort&compress}

\begin{document}


\begin{frontmatter}

\title{Study of the excited $1^-$ charm and charm-strange mesons}

\author[1]{Qiang Li}\ead{lrhit@protonmail.com}
\author[1]{Yue Jiang      \corref{corauthor}}\ead{jiangure@hit.edu.cn}
\author[1]{Tianhong Wang}\ead{thwang@hit.edu.cn}
\cortext[corauthor]{Corresponding author}
\author[1]{Han Yuan}\ead{hanyuan@hit.edu.cn}
\author[1]{  Guo-Li Wang   \corref{corauthor}}\ead{gl\_wang@hit.edu.cn}
\author[2,3]{Chao-Hsi Chang\corref{corauthor}}\ead{zhangzx@itp.ac.cn}

\address[1]{Harbin Institute of Technology, Harbin, 150001, P. R. China}
\address[2]{CCAST (World Laboratory), P.O. Box 8730, Beijing 100080, P. R. China}
\address[3]{Institute of Theoretical Physics, Chinese Academy of Sciences, P.O. Box 2735, Beijing 100080, P. R. China}

\begin{abstract}
We give a systematical study on the recently reported excited charm and charm-strange mesons with potential $1^-$ spin-parity, including the $D^*_{s1}(2700)^+$, $D^*_{s1}(2860)^+$, $D^*(2600)^0$, $D^*(2650)^0$, $D^*_1(2680)^0$ and $D^*_1(2760)^0$. The main strong decay properties are obtained by the framework of Bethe-Salpeter (BS) methods. Our results reveal that the two $1^-$ charm-strange mesons can be well described by the further $2^3\!S_1$-$1^3\!D_1$ mixing scheme with a mixing angle of $8.7^{+3.9}_{-3.2}$ degrees. The predicted decay ratio $\frac{\mathcal{B}(D^*K)}{\mathcal{B}(D~K)}$ for $D^*_{s1}(2860)$ is $0.62^{+0.22}_{-0.12}$.~$D^*(2600)^0$ can also be explained as the $2^3\!S_1$ predominant state with a mixing angle of $-(7.5^{+4.0}_{-3.3})$ degrees. Considering the mass range, $D^*(2650)^0$ and $D^*_1(2680)^0$ are more likely to be the $2^3\!S_1$ predominant states, although the total widths under both the $2^3\!S_1$ and $1^3\!D_1$ assignments have no great conflict with the current experimental data.  The calculated width for LHCb $D^*_1(2760)^0$ seems about 100 \si{MeV} larger than experimental measurement if taking it as $1^3\!D_1$ or $1^3\!D_1$ dominant state $c\bar u$. The comparisons with other calculations and several important decay ratios are also present. For the identification of these $1^-$ charm mesons, further experimental information, such as $\frac{\mathcal{B}(D^{\unseen{*}}\pi)}{\mathcal{B}(D^*\pi)}$ are necessary.
\end{abstract}

\end{frontmatter}


\section{Introduction}
Recently lots of natural parity charm and charm-strange mesons are observed in experiments\,\cite{Belle-2008,BaBar-2009,BaBar-2010,LHCb-2012,LHCb-2013,LHCb-2014,PDG-2014,BaBar-2015,LHCb-2015,LHCb-2016}, which are summarized in \autoref{Tab-Exp}, where we have combined the statistical, systematic and model errors in quadrature for simplicity. These new resonances have great importance in improving our knowledge of the radial and orbital charmed excitations. Especially for the spin-parity $1^-$ charm and charm-strange states, there may exist the $2^3\!S_1$-$1^3\!D_1$ mixing 
, which makes the assignments more complicated.
\begin{table}[!h]
\caption{Experimental results of the recently discovered excited open charm mesons with natural spin-parity.}\label{Tab-Exp}
\vspace{0.2em}\centering
\begin{tabular}{ |p{1.7cm}|p{2.3cm}|p{2.2cm}|p{1.5cm}|p{1.5cm} |p{1.cm}| }
\hline
{Resonance}        				    & Mass~\si{MeV}        &Width~\si{MeV}      &$J^P$       &Refs.                     &Time  \\
\hline
\multirow{5}{*}{$D^{*}_{s1}(2700)^+$}    &$2708\pm14$           &$108\pm36$  & \multirow{5}{*}{$1^-$}  &Belle\cite{Belle-2008}    &2008\\
                                         &$2710\pm12$           &$149\pm52.5$        &            &BaBar\cite{BaBar-2009}    &2009\\
                                         &$2709.3\pm4.9$        &$115.8\pm14.1$      &		  &LHCb\cite{LHCb-2012} 	  &2012\\
                                         &$2709\pm4$            &$117\pm13$          &            &PDG\cite{PDG-2014}        &2014\\
                                         &$2699^{+14}_{-7}$     &$127^{+24}_{-19}$   &            &BaBar\cite{BaBar-2015}    &2015\\
\hline
\multirow{4}{*}{$D^{*}_{sJ}(2860)^+$}    &$2862\pm5.4$          &$48\pm6.7$          &Natural     &BaBar\cite{BaBar-2009}    &2009\\
							    &$2866.1\pm6.4$        &$69.9\pm7.3$        &Natura      &LHCb\cite{LHCb-2012}      &2012\\
 							    &$2859\pm27$		&$159\pm80$     	&$1^-$	  &LHCb\cite{LHCb-2014}	  &2014\\
 							    &$2860.5\pm7$		&$53\pm10$     		&$3^-$	  &LHCb\cite{LHCb-2014}	  &2014\\
\hline		
                    $D^*(2600)^0$        &$2608.7\pm 3.5$       &$93\pm 14.3$        &Natural     &BaBar\cite{BaBar-2010}    &2010 \\
                    $D^*(2650)^0$        &$2649.2\pm 4.9$       &$140\pm 25.5$       &Natural     &LHCb\cite{LHCb-2013}      &2013 \\
                    $D_1^*(2680)^0$      &$2681.1\pm15.1$       &$186.7\pm14.6$      &$1^-$       &LHCb\cite{LHCb-2016}      &2016 \\
                    $D^*(2760)^0$        &$2763.3\pm 3.3$       &$60.9\pm 6.2$       &Natural     &BaBar\cite{BaBar-2010}    &2010 \\
                    $D^*_J(2760)^0$      &$2760.1\pm 3.9$       &$74.4\pm 19.4$      &Natural     &LHCb\cite{LHCb-2013}      &2013 \\
                    $D_3^*(2760)^0$      &$2775.5\pm7.9$        &$95.3\pm 35.4$      &$3^-$       &LHCb\cite{LHCb-2016}      &2016 \\
                    $D_1^*(2760)^0$      &$2781\pm 21.9$        &$177\pm 38.4$       &$1^-$       &LHCb\cite{LHCb-2015}      &2015 \\
                    $D_J^*(3000)^0$      &$3008.1\pm 4.0$       &$110.5\pm 11.5$     & Natural    &LHCb\cite{LHCb-2013}      &2013 \\
                    $D_2^*(3000)^0$      &$3214\pm 56.8$        &$186\pm 81.0$       &$2^+$       &LHCb\cite{LHCb-2016}      &2016 \\
\hline
                    $D^*(2600)^+$        &$2621.3\pm 5.6$       &$93        $        &Natural     &BaBar\cite{BaBar-2010}    &2010 \\
                    $D^*(2760)^+$        &$2769.7\pm 4.1$       &$60.9        $      &Natural     &BaBar\cite{BaBar-2010}    &2010 \\
                    $D^*_J(2760)^+$      &$2771.7\pm4.2$        &$66.7\pm12.4$       & Natural    &LHCb\cite{LHCb-2013}      &2013 \\
                    $D^*_3(2760)^-$      &$2798\pm9.9$          &$105\pm29.8$        & $3^-$      &LHCb\cite{LHCb-2015}      &2015 \\
                    $D_J^*(3000)^+$      &$3008.1$(fixed)       &$110.5$(fixed)      & Natural    &LHCb\cite{LHCb-2013}      &2013 \\
                    \hline
\end{tabular}
\end{table}

$D^*_{s1}(2700)^+$ was first discovered by Belle collaboration in 2008\,\cite{Belle-2008} in channel $D^*_{s1}(2700)^+\to D^0K^+$, and then confirmed by BaBar in 2009\,\cite{BaBar-2009} and LHCb in 2012\,\cite{LHCb-2012}. Furthermore, the BaBar collaboration also obtained two ratios of branching fractions\,\cite{BaBar-2009},
\begin{gather}
R_{K}\left[D^*_{s1}(2700)^+\right] \equiv \frac{\mathcal{B}[D^*_{s1}(2700)^+ \to D^*K]}{\mathcal{B}[D^*_{s1}(2700)^+ \to DK]}=0.91\pm 0.13_{\up{stat}}\pm0.12_{\up{syst}},\label{E-ratio-Dsa}\\
R_{K}[D^*_{sJ}(2860)^+]  \equiv \frac{\mathcal{B}[D^*_{sJ}(2860)^+ \to D^*K]}{\mathcal{B}[D^*_{sJ}(2860)^+ \to DK]}=1.1\pm 0.15_{\up{stat}}\pm0.19_{\up{syst}}\label{E-ratio-Dsb},
\end{gather}
where we have defined the abbreviation $R_K$ for simplicity.
$D^*_{sJ}(2860)^+$ were first detected by BaBar together with the $D^*_{s1}(2700)^+$ and then confirmed by LHCb\,\cite{LHCb-2012}. However, there are about $3\sigma$ discrepancies in the total width. This discrepancy was resolved by LHCb's subsequent measurement with the amplitude analysis in 2014\,\cite{LHCb-2014}, which find that the structure $D^*_{sJ}(2860)^+$ contains both spin-1 and spin-3 components, while a larger width of the former one is preferred. The potential models predict the masses of $2^3\!S_1$ and $1^3\!D_1$ charm-strange mesons are around $2.73$ and $2.90~\si{GeV}$ respectively \cite{God-1985,God-2016}. The $D^*_{s1}(2700)^+$ and $D^*_{s1}(2860)^+$ are then usually interpreted as the $2^3\!S_1$ and $1^3\!D_1$ charge-strange mesons, respectively. There are many works on the properties of these two resonances. The $D^*_{s1}(2700)^+$ is identified as the $2^3\!S_1$ $c\bar s$ in Refs.\,\cite{Col-2008,BC-2009,BC-2011,AZ-2011,Jorge-2015}, while in Refs.\,\cite{God-2014} the $1^3\!D_1$ assignments are favored.
In Refs.\,\cite{DML-2010,XHZ-2010,DML-2011,BC-2011,God2-2014,QTS-2015,BC-2015} the $2S$-$1D$ mixing states of $D^*_{s1}(2700)^+$ and $D^*_{s1}(2860)^+$ are discussed, and we will discuss the mixing scheme in detail in \autoref{Sec-3}. Besides the conventional assignments, Ref.\,\cite{Guo-2011} argued that the $D^*_{sJ}(2860)^+$ can be explained as $D_1(2420)K$ bound states by using the chiral and heavy quark symmetry.

For the corresponding charm mesons, the GI model\,\cite{God-1985,God-2016} predicts the $2^3\!S_1$ and $1^3\!D_1$ states $c\bar u$ locate in the mass range of about $2.64$ and $2.82$ \si{GeV}, respectively, while the mass of $1^3\!D_3$ state is predicted to be quite close to the $1^3\!D_1$ state. BaBar in 2010\,\cite{BaBar-2010} reported two natural parity resonances $D^*(2600)^0$ and $D^*(2760)^0$. Further more, they measured the following ratio of branching fraction as,
\begin{gather}
R_{D^+}\left[D^*(2600)^0\right] \equiv \frac{\mathcal{B}[D^*(2600)^0 \rightarrow D^+\pi^-]}{\mathcal{B}[D^*(2600)^0 \to D^{*+}\pi^-]}=0.32\pm 0.02_{\up{stat}}\pm0.09_{\up{syst}}.\label{E-ratio-Da}
\end{gather}
Again we have introduced an abbreviation $R_{D^+}$ for the sake of simplicity.
Later in 2013 LHCb\,\cite{LHCb-2013} discovered two natural parity charmed particle $D^*(2650)^0$ and $D^*_J(2760)^0$. Then in 2015 LHCb\,\cite{LHCb-2015} reported the $1^-$ state $D^*_1(2760)^0$, which has a large width of $177\pm38$ \si{MeV}. Very recently, in 2016 by using the amplitude analysis, LHCb collaboration measured a $1^-$ state $D^*_1(2680)^0$ and a $3^-$ state $D^*_3(2760)^0$\,\cite{LHCb-2016}. The later one's mass and total width seems consistent with the BaBar $D^*(2760)^0$ and LHCb $D^*_J(2760)^0$. These experimental data are also summarized in \autoref{Tab-Exp}, where the isospin partners of these neutral charm mesons are also listed in the bottom of \autoref{Tab-Exp} for comparison. The $D^*(2760)^0$, $D^*_J(2760)^0$ and $D^*_3(2760)^0$ can be interpreted as the same particle, namely, the $3^-$ state $c\bar u$, while this interpretation is favored by Refs.\,\cite{Zhong-2010,Tian-2016,Zhi-2011,BC-2011}. Then there are still four natural parity resonances, $D^*(2600)^0$, $D^*(2650)^0$, $D^*_1(2680)^0$ and $D^*_1(2760)^0$ in the mass range of $2.6\sims 2.8$ \si{GeV}. In the traditional conventions of charm meson spectroscopy, these four resonances should correspond the $2^3\!S_1$ and $1^3\!D_1$ states $c\bar u$ or the mixtures of them.

These newly observed charm resonances have also been studied with the $2^3\!S_1$, $1^3\!D_1$ assignments or the $2S$-$1D$ mixing scheme in theory by several models, including the non-relativistic quark model\,\cite{Zhong-2010,DML-2011}, the heavy quark effective theory\,\cite{Zhi-2011,Zhi-2013}, the effective Lagrangian approach based on heavy quark chiral symmetry\,\cite{Col-2012}, the EHQ decay formula\,\cite{BC-2011,BC-2015} and the QPC model\,\cite{ZFS-2010,Lv-2014,QTS-2015,QTS2-2015,Yu-2016}. However, the current theoretical calculations for these higher mass charmed mesons can not be well consistent with the experimental data. We find the calculated ratio $R_{D^+}[D^*(2600)^0]$ for taking it as the $2^3\!S_1$ state is usually greater than the experimental value Eq.\,\eref{E-ratio-Da}\,\cite{Zhong-2010,Zhi-2011,Col-2012, Zhi-2013,Lv-2014}, while Ref.\,\cite{Col-2012} argues that no quantum number assignments for pure state at mass $2600$ \si{MeV} is able to reproduce the experimental ratio.

Generally, all the physical mesons have definite $J^P$ spin-parity or $J^{PC}$ for quarkonia. In the relativistic situations, the spin $S$ and orbital angular momentum $L$ are no longer the good quantum numbers, and the physical states are not always located in the definite $^{2S+1}\!L_J$ states. This situations become obvious in the $1^+$ and $1^-$ mesons, for the $1^+$ states we always have to make the $^1\!P_1$-$^3\!P_1$ mixing to fit the physical states\,\cite{Rosner-1986,God-1991}, while for the $1^-$ states the $2^3\!S_1$-$1^3\!D_1$ mixing is needed to fit the experimental measurements\,\cite{Rosner-2005}. So in a more effective and appropriate method to describe the bound state, we should focus on the $J^{P(C)}$, which are the good quantum numbers in any case.
In principal, if we use a full relativistic method to solve the eigenstate problem of the bound mesons with definite $J^{P(C)}$, we do not need mixing to fit the data. We have tried this by BS method in a previous work to study the state $D^*_{s1}(2700)$\,\cite{GLW-2013}, based on the BS wave function constructed directly from the quantum number $J^{P}=1^{-}$. The Salpeter wave functions of $1^-$ states were given, and by solving the full Salpeter equations, we obtain the eigenstates for $c\bar s$ and find that all the states include both $S$ and $D$-wave components. The first state is $1S$ dominant while $D$-wave components can be ignored. The second state is $2S$ dominant, which is the first radial excited state. The third state, which is the second radial excited state, is predominant by $1D$ components.
But our previous results, including the mass spectra and decays can not fit the data very well. The reason is that, we also make some approximations. The first is the instantaneous approximation, which assumes the potential is static, since the four-dimensional BS equation with non-static potential is quite difficult to solve. The second is the interaction kernel, where we choose the Coulomb-like plus linear potential. The Coulomb potential comes from the single-gluon-exchange, where we only keep the first order of QCD interaction. Also the linear confinement potential is introduced by phenomenological analysis. Since BS equation is a integral equation, then the kernel include all the ladder diagrams contributions but not the cross diagrams and the annihilation diagrams.
These approximations have some effects in diagonalizing the mass matrix. So our method is not a full theory and not a full relativistic method, and can not exactly fit the experimental measurements. To overcome this discrepancy, we will make a further mixing to fit the physical states. In this study, we will give a continuous study of these $1^-$ states open charm mesons. We make a further mixing by the second and third radial excited states. Our mixing angle may be smaller than other non-relativistic methods since some relativistic corrections have already been kept in.

In this research, we will calculate the OZI allowed strong decays of these potential $1^-$ charm-strange mesons, $D^*_{s1}(2700)$, $D^*_{s1}(2860)$, and the neutral charm meson $D^*(2600)$, $D^*(2650)$, $D^*_1(2680)$, and $D^*_1(2760)$, where the charge superscripts ``$+$" and ``$0$" are omitted for brevity here and also in the following context. We will focus on the further $2^3\!S_1$-$1^3\!D_1$ mixing scheme to discuss the assignments for these resonances, and the BaBar measured ratios Eqs.~\eref{E-ratio-Dsa} and \eref{E-ratio-Da} are used to restrict the mixing angle.
This work are studied within the framework of the instantaneous Bethe-Salpeter methods\,\cite{BS-1951,Salpeter-1952}.
The BS methods have been widely used and achieved good performance in the strong decays of heavy mesons\,\cite{Chao-2005,ZHW-2012,Tian-2013}, hadronic transition~\cite{JHEP09-2015,QL-2016,QL-2017}, decay constants calculations and annihilation rates~\cite{GLW-2006, PLB674-2009, JHEP03-2013}.

The manuscript is organized as below: In \autoref{Sec-2} we give the theoretical formalisms of the strong decays by BS methods; then in \autoref{Sec-3} the numerical results and detailed discusses are present; finally, we give a brief summary and conclusion about this work.

\section{Theoretic calculations}\label{Sec-2}
In this section first we give a brief review on the calculations of transition matrix element and BS methods; then the $1^-$ states Salpeter wave functions are present.

\subsection{Transition matrix element}
\begin{figure}[h]
\centering
\includegraphics[width=0.55\textwidth]{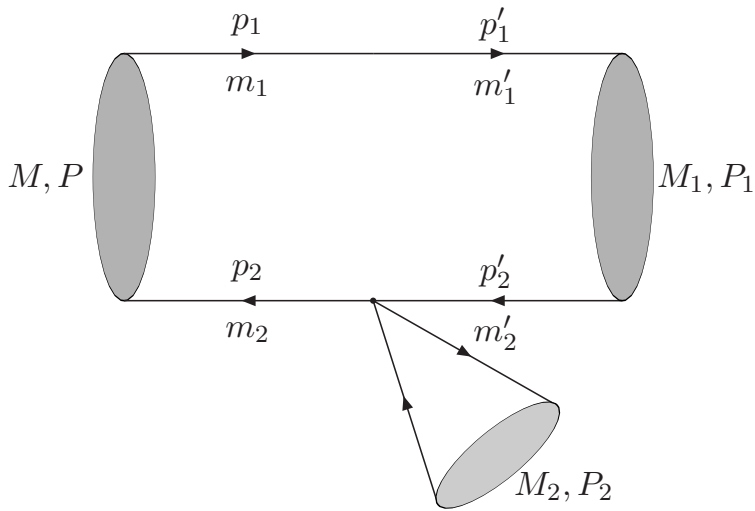}
\caption{Feynman diagram for two body strong decay of $D^*_{(s)}$. $m_1=m_1'=m_c$, is the constituent mass of $c$ quark. $M_2$ and $P_2$ denote the mass and momentum of the final light meson, respectively.}\label{Fig-Feynman}
\end{figure}
The Feynman diagram for strong decays of charmed meson is showed in \autoref{Fig-Feynman}, where we use subscript 1 and 2 to denote the final charmed meson and light meson, respectively. By using the reduction formula, the transition matrix element for decay $D^*_s \to D^{(*)}K$ can be written as\,\cite{Chao-2005},
\begin{gather}\label{E-matrix1}
\langle D^{(*)}(P_1)K(P_2)|D^*_s(P) \rangle = \int \up{d}^4 x e^{-iP_2\cdot x}(M_2^2-P^2_2) \langle D^{(*)}(P_1)|\varPhi_2(x)|D^*_s(P) \rangle,
\end{gather}
where $P$, $P_1$ and $P_2$ denote the momenta of initial state $D^*_s$, final charmed meson $D^{(*)}$ and final light meson $K$, respectively~(see \autoref{Fig-Feynman}); $M_2$ is the mass of final light meson. $\varPhi_2(x)$ is used to describe the light scalar meson field. The PCAC relation reads
\begin{gather}
\varPhi_2(x)=\frac{1}{M_2^2F_2} \partial_\mu(\bar s \Gamma^\mu q),
\end{gather}
where $F_2$ is the decay constant of the light scalar meson; $q=u$ or $d$ corresponds to the $K^+$ and $K^0$ respectively; the abbreviation $\Gamma^\mu=\gamma^\mu \gamma^5$ is used. Inserting the PCAC relation into Eq.~(\ref{E-matrix1}), with the low energy theorem, Eq.~\eref{E-matrix1} can be expressed as
\begin{gather}\label{E-matrix2}
\langle D^{(*)}(P_1)K(P_2)|D^*_s(P) \rangle =(2\pi)^4 \delta^4(P-P_1-P_2) \frac{-i P_2^\mu}{F_2} \langle D^{(*)}(P_1)|\bar s \Gamma_\mu q|D^*_s(P) \rangle.
\end{gather}
Then the decay amplitude $\mathcal{M}$ can be described as,
\begin{gather}
\mathcal{M}=\frac{P_2^\mu}{F_2}\langle D^{(*)}(P_1)|\bar s \Gamma_\mu q|D^*_s(P) \rangle,
\end{gather}
where the transition matrix element $\langle D^{(*)}(P_1)|\bar s \Gamma_\mu q|D^*_s(P) \rangle$ can be calculated by Salpeter method and will be derived in next subsection. The decay width $\Gamma$ is then expressed as,
\begin{gather}
\Gamma=\frac{1}{8\pi} \langle|\mathcal{M}|^2\rangle \frac{|\vec{P_1}|}{M^2},
\end{gather}
where $|\vec{P}_1|=\frac{1}{2M}\sqrt{\lambda(M,M_1,M_2)}$ and the K\"all\'en function $\lambda(a,b,c)=(a^2+b^2+c^2-2ab-2bc-2ac)$ is used; $\langle|\mathcal{M}|^2\rangle$ stands for the average over initial spins and sum over final spins;

When the light meson is $\eta$, the $\eta-\eta'$ mixing should be considered. In this work we use the following mixing conventions,
\begin{equation}\label{E-mix-eta}
\begin{bmatrix}\eta\\ \eta' \end{bmatrix}= \begin{bmatrix}\cos\theta_\eta & \sin\theta_\eta \\ -\sin\theta_\eta& \cos\theta_\eta \end{bmatrix} \begin{bmatrix}\eta_8\\ \eta_1 \end{bmatrix}.
\end{equation}
$\eta_8$ and $\eta_1$ are the SU(3) octet and singlet states, respectively. We use the mixing angle $\theta_\eta=19\degree$.
To include this mixing effect, the PCAC relation reads
\begin{align}
\varPhi_\eta(x)&=\cos\theta_\eta \varPhi_{\eta_8}(x) + \sin\theta_\eta \varPhi_{\eta_1}(x)\notag \\
               &= \frac{\cos\theta_\eta}{M_{\eta_8}^2 f_{\eta_8}}\partial_\mu \left ( \frac{\bar u \Gamma^\mu u + \bar d \Gamma^\mu  d -2\bar s \Gamma^\mu  s} {\sqrt{6}} \right)+ \frac{\sin \theta_\eta}{M_{\eta_1}^2 f_{\eta_1}}\partial_\mu \left ( \frac{\bar u \Gamma^\mu  u + \bar d \Gamma^\mu d +\bar s \Gamma^\mu s} {\sqrt{3}} \right)\notag \\
               &=\left[ \frac{-2\cos\theta_\eta}{\sqrt{6}M^2_{\eta_8}f_{\eta_8}} + \frac{\sin\theta_\eta}{\sqrt{3}M^2_{\eta_0}f_{\eta_0}} \right]\partial_\mu (\bar s \Gamma^\mu s)
\end{align}
where in the last step, we have only remained the $\bar s\Gamma^\mu  s$ part since others have no contribution here; $f_{\eta_8}$ and $f_{\eta_1}$ are the corresponding decay constants of $\eta_8$ and $\eta_1$, respectively.

When the $\pi^0$ is involved in the final states, the PCAC relation reads
\begin{align}
\varPhi_{\pi^0}(x)&=\frac{1}{M_{\pi^0}^2f_{\pi}}\partial_\mu \left(\frac{\bar u \Gamma^\mu  u - \bar d \Gamma^\mu  d}{\sqrt{2}}\right)\notag\\
               &=\frac{1}{\sqrt{2} M_{\pi^0}^2 f_{\pi}}\partial_\mu \left( \bar u \Gamma^\mu  u \right)
\end{align}
Again we have only kept the contributory part.

\subsection{Transition matrix element with Salpeter wave function}
In this subsection we briefly review the BS methods. The BS equation is an four-dimensional integral equation, which reads in momentum space as~\cite{BS-1951}
\begin{gather}\label{BS}
(\slashed p_1 - m_1)\Psi(q)(\slashed p_2+m_2)=\si{i}\int \frac{\text{d}^4k}{(2\uppi)^4}V(q-k)\Psi(k),
\end{gather}
where $\Psi(q)$ is the four dimensional BS wave function; $V(q-k)$ stands for the BS interaction kernel; $p_1$ and $p_2$ are the quark and anti-quark momentum respectively, while $m_1$ and $m_2$ are the corresponding masses~(see \autoref{Fig-Feynman}). It is more convenient to express the $p_1$ and $p_2$ with the total momentum $P$ and inner relative momentum $q$ as
\begin{equation}
	p_1=\alpha_1P+q,    \qquad p_2=\alpha_2 P - q.
\end{equation}
$\alpha_i~(i=1,2)$ is defined as $\alpha_i \equiv \frac{m_i}{m_1+m_2}$.
Salpeter wave function $\varphi(q_\perp)$ is related to BS wave function $\Psi(q)$ by the following definition
\begin{gather}
\varphi(q_\perp) \equiv \text{i}\int \frac{\mathrm{d}q_P}{2\pi}\Psi(q),\quad \eta(q_\perp) \equiv \int \frac{\mathrm{d}^3k_\perp}{(2\pi)^3}\varphi(k_\perp)V(|q_\perp- k_\perp|),
\end{gather}
where $q_P=\frac{P\cdot q}{M}$ and $q_\perp=q-\frac{P}{M}q_P$, in rest frame of initial meson they correspond to the $q^0$ and $\vec q$ respectively; the 3-dimensional integration $\eta(q_\perp)$ can be understood as the BS vertex for bound states; $V(|q_\perp- k_\perp|)$ denotes the instantaneous interaction kernel, namely, the inner interaction are assumed to be a static potential. As usual, in this work, the specific interaction kernel $V(r)$ we use are the Coulomb-like potential plus the unquenched scalar confinement one\,\cite{GLW-2006},
\begin{gather}
V(r)=V_s(r)+V_0+\gamma_0 \otimes \gamma^0 V_v(r)= \frac{\lambda}{\alpha}(1-e^{-\alpha r})+V_0-\frac{4}{3}\frac{\alpha_s}{r}e^{-\alpha r},
\end{gather}
where $\lambda$ is the string constant, $\alpha_s(r)$ is the running strong coupling constant, and $V_0$ is a free constant fixed by fitting the data.
By Fourier transformation, the potential $V(\vec q\,)$ in momentum space reads,
\begin{gather}
V(\vec q\,)=-\left(\frac{\lambda}{\alpha}+V_0\right)(2\pi)^3 \delta^3(\vec q\,)+\frac{\lambda}{\pi^2}\frac{1}{(\vec q\,^2+\alpha^2)^2}-\frac{2}{3\pi^2}\frac{\alpha_s(\vec q\,)}{(\vec q\,^2+\alpha^2)},
\end{gather}
where the running coupling constant $\alpha_s(\vec q\,)=\frac{12\pi}{27}\frac{1}{\log(a+\vec q\,^2/\Lambda^2_\up{QCD})}$.

Then under the instantaneous approximation, the BS equation (\ref{BS}) can be written as
\begin{gather}\label{BS-3D}
\Psi(q)=S(p_1)\eta(q_\perp)S(-p_2).
\end{gather}
$S(p_1)$ and $S(-p_2)$ are the propagators for the quark and anti-quark respectively, and can be decomposed as
\begin{equation}\label{Si}
\begin{aligned}
S(+p_1)&=\frac{\si{i}\Lambda_1^+}{q_P+\alpha_1M-\omega_1+\si{i}\epsilon}+\frac{\si{i}\Lambda_1^-}{q_P+\alpha_1M+\omega_1-\si{i}\epsilon},\\
S(-p_2)&=\frac{\si{i}\Lambda_2^+}{q_P-\alpha_2M+\omega_2-\si{i}\epsilon}+\frac{\si{i}\Lambda_2^-}{q_P+\alpha_2M-\omega_2+\si{i}\epsilon},
\end{aligned}
\end{equation}
where $\omega_i=\sqrt{m_i^2-q^2_\perp}~(i=1, 2)$. $\Lambda^{\pm}_i(q_\perp)$~($i=1,2$) are the projection operators, which have the following forms,
\begin{gather}
\Lambda^{\pm}_i(q_\perp)=\frac{1}{2\omega_i}\left[ \frac{\slashed P}{M}\omega_i\pm(-1)^{i+1}(m_i+\slashed q_\perp) \right].
\end{gather}
It can be easily check that, the projection operators satisfy the following relations:
\begin{gather}\label{E-Lambda}
\Lambda^+_i(q_\perp) + \Lambda^-_i(q_\perp)=\frac{\slashed P}{M},\quad
\Lambda^{\pm}_i(q_\perp) \frac{\slashed P}{M} \Lambda^\pm_i(q_\perp)=\Lambda^{\pm}_i(q_\perp), \quad
\Lambda^{\pm}_i(q_\perp) \frac{\slashed P}{M} \Lambda^\mp_i(q_\perp)=0.
\end{gather}

Since the BS kernel is assumed to be instantaneous, we can perform a contour integration over $q_P$ on both sides of Eq.~(\ref{BS-3D}), then we achieve the Salpeter equation as
\begin{gather} \label{E-Salpeter}
\varphi(q_\perp)=\frac{\Lambda_1^+(q_\perp)\eta(q_\perp)\Lambda^+_2(q_\perp)}{(M-\omega_1-\omega_2)} - \frac{\Lambda_1^-(q_\perp)\eta(q_\perp)\Lambda^-_2(q_\perp)}{(M+\omega_1+\omega_2)}.
\end{gather}
To make further simplification, we introduce four new wave functions $\varphi^{\pm\pm}(q_\perp)$ with the definitions as
\begin{gather}\label{E-Phis}
\varphi^{\pm\pm}(q_\perp)=\Lambda^{\pm}_i(q_\perp) \frac{\slashed P}{M}\varphi(q_\perp)\frac{\slashed P}{M} \Lambda^\pm_i(q_\perp),
\end{gather}
where $\varphi^{++}$ is then called the positive Salpeter wave function, while $\varphi^{--}$ is called the negative Salpeter wave function.

Then with the help of Eqs.\,(\ref{E-Lambda}), the Salpeter equation (\ref{E-Salpeter}) can be further expressed as the following 4 coupled equations~\cite{Salpeter-1952}
\begin{gather}
\varphi^{+-}(q_\perp)=\varphi^{-+}(q_\perp)=0\label{BS-np},\\
(M-\omega_1-\omega_2)\varphi^{++}(q_\perp)=+\Lambda_1^+(q_\perp)\eta(q_\perp)\Lambda^+_2(q_\perp),\label{BS-pp}\\
(M+\omega_1+\omega_2)\varphi^{--}(q_\perp)=-\Lambda_1^-(q_\perp)\eta(q_\perp)\Lambda^-_2(q_\perp).\label{BS-nn}
\end{gather}
From above equations, we can see that in the weak binding condition, namely, $M\sim (\omega_1+\omega_2)$, $\varphi^{--}$ is much smaller compared with $\varphi^{++}$ and can be ignored in the calculations. However, these four equations play equivalent roles in solving the eigenstate problem. The normalization condition for Salpeter wave function reads
\begin{gather}\label{Norm-BS}
\int \frac{\text{d}^3q_\perp}{(2\uppi)^3}\left[\overline\varphi^{++}\frac{\slashed P}{M}\varphi^{++}\frac{\slashed P}{M}-\overline\varphi^{--}\frac{\slashed P}{M}\varphi^{--}\frac{\slashed P}{M}\right]=2M.
\end{gather}

According to Mandelstam formalism\,\cite{Man-1955}, the transition matrix element $\langle D^{(*)}(P_1)|\bar s \Gamma_\mu q|D^*_s(P) \rangle$ can be expressed as
\begin{gather}\label{E-TMatrix}
\langle D^{(*)}(P_1)|\bar s \Gamma^\mu q|D^*_s(P) \rangle \simeq \int \frac{\up{d}^3 \vec{q}}{(2\pi)^3}\up{Tr} \left[ \bar \varphi^{\prime++}_{P_1}(|\vec{q}\,'|) \frac{\slashed P}{M} \varphi^{++}_P(|\vec{q}\,|) \Gamma^\mu \right],
\end{gather}
where $\bar \varphi^{\prime++}_{P_1}$ is defined as $\gamma^0( \varphi^{\prime++}_{P_1})^\dagger \gamma^0$, and $\varphi^{\prime++}_{P_1}$ is the positive Salpeter wave function of the final state; $\vec q\,'=\vec q-\frac{m_1'}{m_1'+m_2'}\vec P_1$; $m_1'$ and $m_2'$ are the constituent quark and anti-quark masses in the final charmed meson~(see \autoref{Fig-Feynman}). To achieve a final result, we still need to know the specific form of the corresponding Salpeter wave functions.

\subsection{Salpeter wave function}\label{Sec-3}
The Salpeter wave functions involved in this calculations include the $0^-$, $1^{-}$ and $1^+$ states, which corresponds to the $^1\!S_0$, $^3\!S_1(^3\!D_1)$ and $^1\!P_1(^3\!P_1)$ states within the non-relativistic models. The Salpeter wave function of $0^-$ state can be seen in Ref.\,\cite{Kim-2004}. Here we only give the $1^-$ state Salpeter wave function. The $^3\!S_1$ and $^3\!D_1$ states share the same spin-parity $J^{P}=1^{-}$. We rewrite the Salpeter wave function for $1^{-}$ states\,\cite{GLW-2006} as below,
\begin{gather}\label{E-1-wave}
\varphi(1^{-})= \frac{q_\perp \!\cdot\! \xi}{|\vec q\,|} \left (f_1  + f_2  \frac{\slashed P}{M} + f_3  \frac{\slashed q_\perp}{|\vec q\,|} + f_4 \frac{\slashed P \slashed q_\perp}{M|\vec q\,|} \right)+ i\frac{\epsilon_{\mu P q_\perp \xi}}{M|\vec q\,|}\gamma^\mu \left(f_5  \frac{\slashed P \slashed q_\perp}{M|\vec q\,|}+  f_6 \frac{\slashed q_\perp}{|\vec q\,|} + f_7  \frac{\slashed P}{M}+ f_8  \right) \gamma^5,
\end{gather}
where $f_i~(i=1,2,\cdots,8)$ are the radial wave functions; $\epsilon_{\mu P q_\perp \xi}=\epsilon_{\mu \nu \alpha \beta }{P^\nu q^\alpha_\perp \xi^\beta}$ and $\epsilon_{\mu \nu \alpha \beta}$ is the totally antisymmetric Levi-Civita tensor; $\xi$ denotes the polarization vector for initial state and fulfills $P\cdot \xi=0,~\sum \xi_\mu^{(r)}\xi_\nu^{(r)}=\frac{P_\mu P_\nu}{M^2}-g_{\mu \nu}$. By using Salpeter equations (\ref{BS-np}), we obtain the following 4 constraint conditions,
\begin{equation}
\begin{aligned}
 f_1  &  =-\frac{|\vec q\,|(\omega_1+\omega_2)}{m_1\omega_2+m_2\omega_1}f_3,  & \quad f_7  &  =\frac{|\vec q\,|(\w_1-\omega_2)}{m_1\omega_2+m_2\omega_1}f_5,      \\
 f_2  &  =-\frac{|\vec q\,|(\omega_1-\omega_2)}{m_1\omega_2+m_2\omega_1}f_4,  & \quad f_8  &  =\frac{|\vec q\,|(\w_1+\omega_2)}{m_1\omega_2+m_2\omega_1}f_6.
 \end{aligned}
\end{equation}
which left us 4 independent wave functions $f_3,~f_4,~f_5$ and $f_6$, only depending on $|\vec{q}\,|$ directly.
In the paper, $\w_i$ is defined as $\sqrt{m_i^2+\vec q\,^2}$ ($i=1,2$).
It can be easily check that, with above Salpeter wave function form, every item in Eq.~\eref{E-1-wave} has the same quantum number $J^P=1^-$. Noticed that this wave function form and constraint conditions for $1^-$ state are not exactly the same with that in Ref.\,\cite{GLW-2006}, however, it can be proved that the two forms are totally equivalent.

According to the definitions Eq.\,(\ref{E-Phis}), the positive Salpter wave function $\varphi^{++}$ for $1^-$ state is then expressed as
\begin{gather}\label{E-1-Phi++}
\varphi^{++}(1^{-})= \frac{q_\perp \!\cdot\! \xi}{|\vec q\,|} \left (A_1  + A_2  \frac{\slashed P}{M} + A_3  \frac{\slashed q_\perp}{|\vec q\,|} + A_4 \frac{\slashed P \slashed q_\perp}{M|\vec q\,|} \right)+ i\frac{\epsilon_{\mu P q_\perp \xi}}{M|\vec q\,|}\gamma^\mu \left(A_5 + A_6 \frac{\slashed P}{M} +  A_7 \frac{\slashed q_\perp}{|\vec q\,|}+ A_8 \frac{\slashed P \slashed q_\perp}{M|\vec q\,|} \right) \gamma^5.
\end{gather}
And the corresponding coefficients $A_i$ are
\begin{equation}
\begin{aligned}
A_1 &=\frac{-q(\w_1+\w_2)}{(m_1\omega_2+m_2\omega_1)}A_3 ,\\
A_2 &=\frac{-q(\w_1-\w_2)}{(m_1\omega_2+m_2\omega_1)}A_4 ,\\
A_3 &=\frac{1}{2}\left(f_3 +\frac{m_1+m_2}{\w_1+\w_2}f_4   \right),\\
A_4 &=\frac{1}{2}\left(f_4 +\frac{\w_1+\w_2}{m_1+m_2}f_3   \right),
\end{aligned}
\qquad \qquad
\begin{aligned}
A_5 &=\frac{q(\w_1+\w_2)}{m_1\omega_2+m_2\omega_1}A_7 ,\\
A_6 &=\frac{q(\w_1-\w_2)}{m_1\omega_2+m_2\omega_1}A_8 ,\\
A_7 &=\frac{1}{2}\left(f_6 -\frac{m_1+m_2}{\omega_1+\omega_2}f_5 \right),\\
A_8 &=\frac{1}{2}\left(f_5 -\frac{\omega_1+\omega_2}{m_1+m_2}f_6 \right).
\end{aligned}
\end{equation}
The negative Salpeter wave function $\varphi^{--}$ can be obtained similarly or by $\varphi^{--}=\varphi-\varphi^{++}$. Then inserting the expressions of $\varphi^{++}$ and $\varphi^{--}$ into the coupled Salpeter equations (\ref{BS-pp}) and (\ref{BS-nn}), we achieve the radial eigenvalue equations, which can be solved numerically. The normalization condition for $1^-$ states Salpeter wave functions now becomes,
\begin{gather}
\int \frac{\up{d}\vec q}{(2\pi)^3} \frac{8\w_1\w_2}{3M(m_1\w_2+m_2\w_1)}(f_3f_4-2f_5f_6)=1.
\end{gather}
Interested readers can see a more detailed procedures on solving the full Salpeter equations in our previous works\,\cite{QL-2016,Kim-2004,Tian2-2013,Tian2-2016}.

\begin{figure}[ht]
\vspace{0.5em}
\centering
\subfigure[BS radial wave functions for the first radial excited state of $D^*_{s1}$.]{\includegraphics[width=0.48\textwidth]{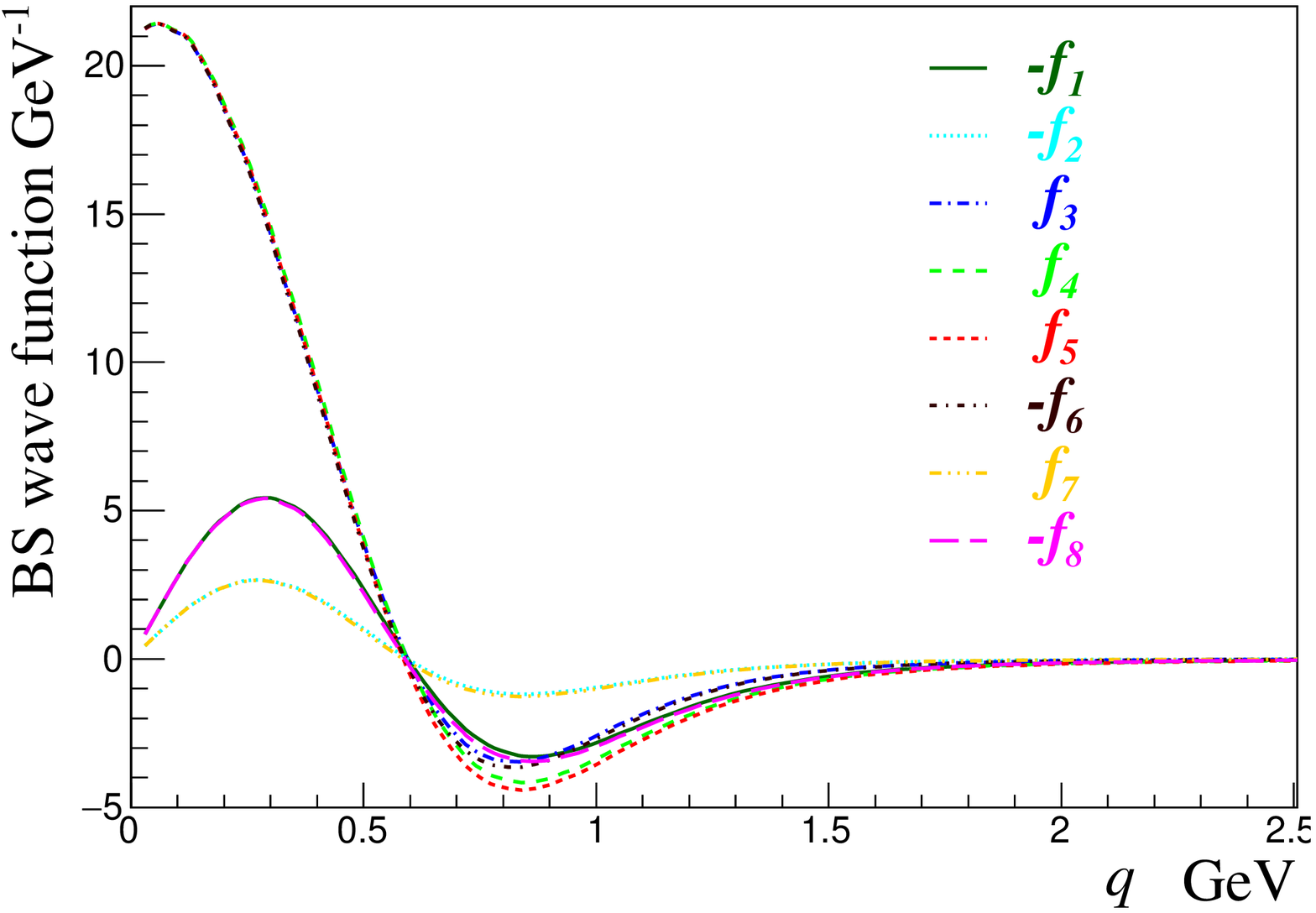} \label{Fig-cs-2S}}
\subfigure[BS radial wave functions for the second radial excited state of $D^*_{s1}$.]{\includegraphics[width=0.48\textwidth]{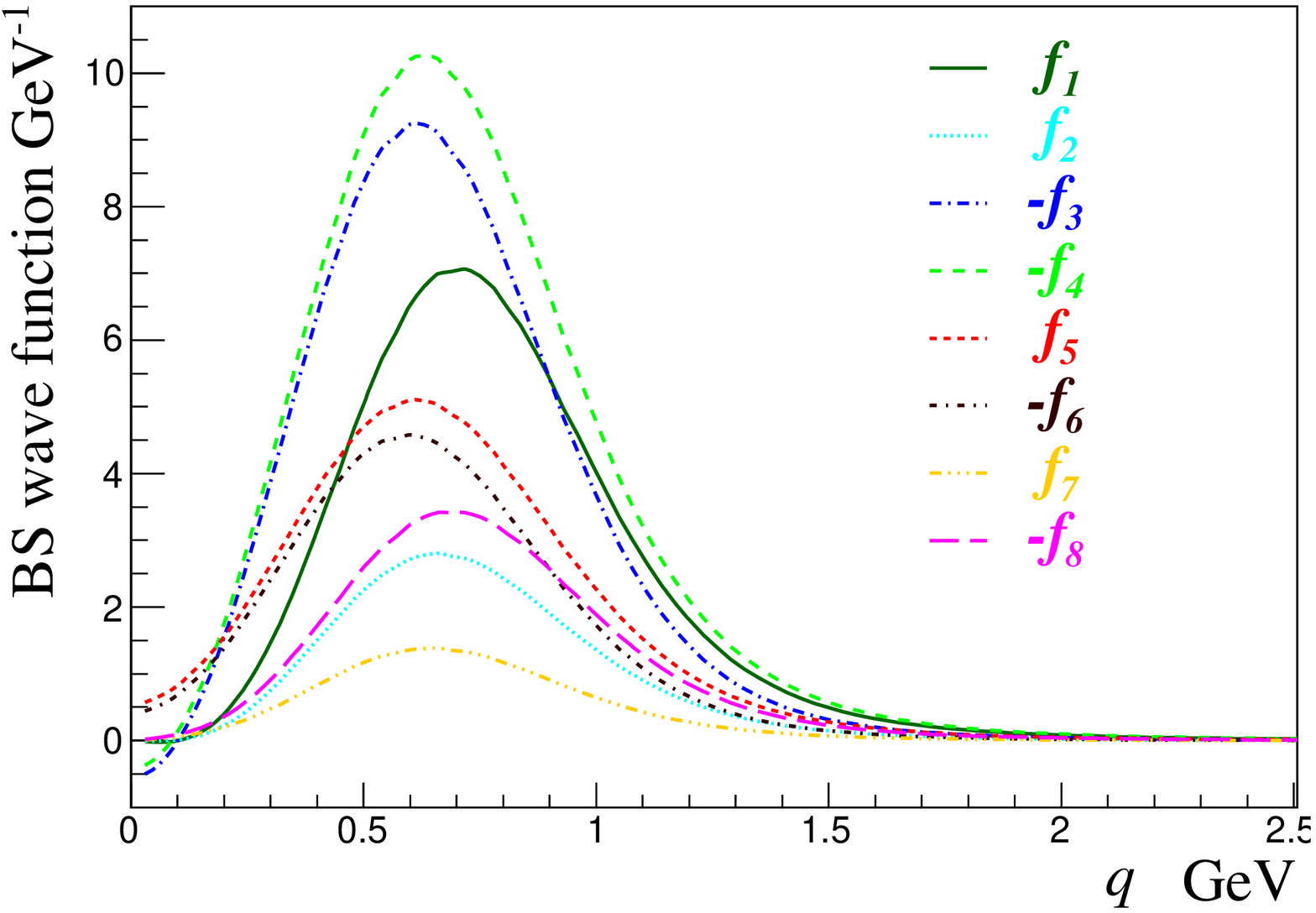} \label{Fig-cs-1D}}\\
\caption{BS wave function for $1^-$ radial excited states of $D^*_{s1}$ mesons.}\label{Fig-wave}
\vspace{0.5em}
\end{figure}

By solving the Salpeter equations, finally we achieve these eight radial wave functions numerically, which are showed in \autoref{Fig-wave}. \autoref{Fig-cs-2S} shows the 8 radial wave functions of the first radial excited state, and \autoref{Fig-cs-1D} shows the radial wave functions of the second radial excited state. 
From the two diagrams, also considering that in Eq.~\eref{E-1-wave}, the direction of momentum $\slashed q_\perp$ has contribution to the $S$ or $D$ wave\,\cite{Chang-2005}, we can conclude that both the first and second radial excited states have $S$ and $D$ wave components, while the first radial excited state is $2^3\!S_1$ predominant and the second radial excited state is a $1^3\!D_1$ dominant state as has been stated in Ref.\,\cite{ChangWang-2010}.

Since the decay final states include $1^+$ meson, we show our treatment of their wave functions.
Generally, the bound state mesons consisting of unequal masses of quark and anti-quark do not have definite charge conjugation parity. So the physical $1^+$ states $D_1$ and $D'_{1}$ can be considered as the admixtures of $1^{++}~(^3\!P_1)$ and $1^{+-}~({^1\!P_1})$ states.
Here we will follow the mixing conventions in Refs.~\cite{Matsuki-2010,Fau-2010}, where the mixing form for $1^+$ states is defined by the mixing angle $\alpha_{1P}$ as
\begin{equation}\label{def-D2}
\begin{bmatrix}|D_1\rangle\\ |D'_1\rangle \end{bmatrix}= \begin{bmatrix}\cos\alpha_{1P} & \sin\alpha_{1P} \\ -\sin\alpha_{1P}& \cos\alpha_{1P} \end{bmatrix} \begin{bmatrix}|1^{+-}\rangle \\ |1^{++}\rangle \end{bmatrix}.
\end{equation}
The heavy quark effective predicts that, in the limit $m_Q\to \infty$ the mixing angle for $1^+$ states are expressed as $\alpha_{1P}=\arctan\sqrt{1/2}=35.3\degree$. This result will be used in the strong decay calculations when $D_1^{(\prime)}$ mesons are involved in the final states. The Salpeter wave functions for $1^{+-}$ and $1^{++}$ states can be found in Ref.\,\cite{Guo-2007}.

Having these numerical Salpeter wave functions, we can calculate the 3-dimensional integral of the transition matrix element $\langle D^{(*)}(P_1)|\bar s \Gamma^\mu q|D^*_s(P) \rangle$ in Eq.\,(\ref{E-TMatrix}). The detailed information on performing this integral can be found in our previous work Refs.\,\cite{Tian-2013, QL-2017}.

\section{Numerical Results and Discussions}\label{Sec-3}
First we specify the corresponding parameters used in this work. The constituent quark masses and other parameters to characterize the model are before\,\cite{Tian-2013}
\begin{align*}
a&=e=2.7183   	   				&\alpha &=0.060~\si{GeV}, &\lambda&=0.210~\si{GeV}^2\\
\Lambda_\text{QCD}&=0.270~\si{GeV}  	&m_u&=0.305~\si{GeV},     &m_d    &=0.311~\si{GeV},\\
 m_s&=0.500~\si{GeV},      		&m_c&=1.620~\si{GeV},     &m_b    &=4.960~\si{GeV}.
\end{align*}
The free parameter $V_0$ is fixed by fitting the mass eigenvalue to experimental value.
The decay constants we used are $f_\pi=130.4$ \si{MeV}, $f_K=156$ \si{MeV}\,\cite{PDG-2014}, $f_{\eta_8}=1.26f_\pi$ and $f_{\eta_1}=1.07f_\pi$. The mixing angle $\theta$ between $\eta-\eta'$ we choose is $\theta_\eta=19\degree$ with the mixing convention in Eq.~\eref{E-mix-eta}. Other involved parameters are from PDG data\,\cite{PDG-2014} unless otherwise specified.

\subsection{$1^-$ charm-strange mesons}
Both the $D^*_{s1}(2700)$ and $D^*_{s1}(2860)$ share spin-parity $J^P=1^-$ determined by experiments. In the first place, we take them as the pure first and second radial excited states, which are dominant by $2^3\!S_1$ and $1^3\!D_1$ respectively. So in this work we still label the first radial excited state as $2^3\!S_1$ and the second excited state as $1^3\!D_1$. The calculated main strong decays properties are listed in \autoref{Tab-Ds-width1}.
\begin{table}[ht]
\caption{Decay widths of $D^*_{s1}(2710)$ and $D^*_{s1}(2860)$ as the $2{^3\!S_1}$ and $1{^3\!D_1}$ dominant $c\bar s$ states in $\si{MeV}$.}\label{Tab-Ds-width1}
\vspace{0.2em}\centering
\begin{tabular}{ |l|c|c|c|c| }
\hline
\multirow{2}{*}{Mode}       &\multicolumn{2}{c|}{$2\,^3\!S_1$}		& \multicolumn{2}{c|}{$1^3\!D_1$}  \\                                                           
\cline{2-5}
		        &$D^*_{s1}(2700)$		   &$D^*_{s1}(2860)$	&$D^*_{s1}(2700)$	&$D^*_{s1}(2860)$  \\
\hline
$D^{*0}K^+$       & 27.5				 &50.0			&6.4  			&13.0\\
$D^{0}K^+$        & 17.8                    &20.8			&28.1  			&38.4\\
$D^{*+}K^0$       & 26.6                    &49.8			&6.1  			&12.7\\
$D^{+}K^0$        & 17.8                    &21.3			&7.5  			&38.4\\
$D^{*+}_{s}\eta$  & 0.9                     &7.6			&0.1  			&1.3\\
$D^{+}_{s}\eta$   & 2.8                     &6.0			&3.1  			&7.2\\
\hline
Total		        &93.4                     &155.5			&51.3  			&111\\
\hline
$\frac{\mathcal{B}(D^*\!K)}{\mathcal{B}(D^{\unseen{*}}\!K)}$
                   &1.52                     &2.37			&0.35 			&0.33\\
\hline
\end{tabular}
\end{table}
From \autoref{Tab-Ds-width1} we can see that, for $D^*_{s1}(2700)$, neither the $2^3\!S_1$ nor $1^3\!D_1$ assignment can produce the experimental ratio $R_K[D^*_{s1}(2700)]=0.91$ though the $2^3\!S_1$ and $1^3\!D_1$ assignments to $D^*_{s1}(2700)$ and $D^*_{s1}(2860)$ respectively are roughly consistent with the experimental measurements. Also notice that the total width is only the half of experimental value when taking $D^*_{s1}(2700)$ as the $1^3\!D_1$ state. So these assignments can not produce experimental data. One also notes that the predicted total decay widths in this paper are much larger than our previous calculation \cite{GLW-2013}, the reason is that we have chosen different $D^*_{s1}(2700)$ mass as input, besides the difference of phase space, the node structure of $2S$ state also has sensitive effect due to the variance of phase space.
%
%

Then we introduce the further $2^3\!S_1$-$1^3\!D_1$ mixing scheme in the $1^-$ charm-strange system. The mixing form we used is defined as,
\begin{equation}
\begin{bmatrix}|D^{*}_{sa}\rangle\\|D^*_{sb} \rangle  \end{bmatrix}= \begin{bmatrix}\cos\theta_s & \sin\theta_s \\ -\sin\theta_s& \cos\theta_s \end{bmatrix} \begin{bmatrix}|2^3\!S_1\rangle \\ |1^3\!D_1\rangle \end{bmatrix}.
\end{equation}
and the mixing states $D^*_{sa}$ and $D^*_{sb}$ correspond to $D^*_{s1}(2700)$ and $D^*_{s1}(2860)$, respectively. To get the experimental branching ratio $R_K[D^*_{s1}(2700)]=0.91^{-0.18}_{+0.18}$, we obtain the mixing angle $\theta_s=(8.7^{+3.9}_{-3.2})\degree$. The errors in experimental ratio $R_K[D^*_{s1}(2700)]$ are combined in quadrature for simplicity. The uncertainties in our mixing angle $\theta_s$ and other obtained results are induced by varying the experimental ratio $R_K[D^*_{s1}(2700)]$ in $1\sigma$ range.

The decay properties with $2S$-$1D$ mixing are listed in \autoref{Tab-Ds-width2}. The total width for $D^*_{s1}(2700)$ is about $100.8$ \si{MeV}, which agrees well with the experimental measurement $\Gamma=117\pm13$~\si{MeV}\,\cite{PDG-2014}. Our results are also consistent with that in Ref.\,\cite{QTS-2015}, where the mixing angle is about $6.8\degree\sims11.2\degree$ and the calculated $\Gamma[D^*_{s1}(2700)]$ is about 100 \si{MeV}. With the obtained mixing angle, the total width for $D^*_{s1}(2860)$ we obtain is $108.8$ \si{MeV}, which is also comparable with the LHCb result $\Gamma[D^*_s(2860)]=159\pm 80.3$ \si{MeV}, but less than the result $\sims 300$ \si{MeV} in Ref.\,\cite{QTS-2015}. Furthermore, the predicted ratio $R_K[D^*_{s1}(2860)]=0.62$, which is also consistent with the result $0.6\sims0.8$ in Refs.\,\cite{QTS-2015,Jorge-2015,BC-2015} and could be used to test this $2^3\!S_1$-$1^3\!D_1$ mixing scheme in the future measurements. The comparisons of our results with other predictions can be seen in \autoref{Tab-Ds-w2}.
\begin{table}[ht]
\caption{Decay widths in $\si{MeV}$ for $D^*_{s1}(2700)$ and $D^*_{s1}(2860$) under the further $2{^3\!S_1}$-$1{^3\!D_1}$ mixing. The mixing angle $\theta_s$ is in unit of degree.}\label{Tab-Ds-width2}
\vspace{0.2em}\centering
\begin{tabular}{ |l|c|c|c|c| }
\hline
$\theta_s$       & \multicolumn{2}{c|}{$8.7^{+3.9}_{-3.2}$}  &  \multicolumn{2}{c|}{$-(76.9^{+2.2}_{-1.8})$}  \\
\cline{1-5}
Mode              & $D^{*+}_{s1}(2700)$   & $D^{*+}_{s1}(2860)$ 	& $D^{*+}_{s1}(2700)$   & $D^{*+}_{s1}(2860)$\\

\hline
$D^{*0}K^+$       & $23.3^{-2.1}_{+1.5}$   	&$19.6^{+3.2}_{-2.5}$   		&$14.9^{-1.5}_{+1.3}$		&$36.1^{+2.7}_{-2.1}$	 \\
$D^{0}K^+$        & $25.2^{+3.5}_{-2.8}$   	&$31.2^{-3.6}_{+2.7}$   		&$16.2^{+1.9}_{-1.4}$		&$32.7^{-1.8}_{+1.6}$	 \\
$D^{*+}K^0$       & $22.5^{-1.9}_{+1.5}$   	&$19.3^{+3.2}_{-2.5}$   		&$14.3^{-1.5}_{+1.3}$		&$36.1^{+2.6}_{-2.1}$	 \\
$D^{+}K^0$        & $25.2^{+3.3}_{-2.8}$  		&$31.1^{-3.6}_{+2.8}$   		&$15.9^{+1.8}_{-1.5}$		&$33.3^{-1.8}_{+1.6}$	 \\
$D^{*+}_{s}\eta$  & $0.8^{-0.1}_{+0.0}$    	&$2.1^{+0.4}_{-0.3}$    		&$0.4^{-0.1}_{+0.0}$		&$5.7^{+0.4}_{-0.2}$	\\
$D^{+}_{s}\eta$   & $3.8^{+0.4}_{-0.4}$     	&$5.5^{-0.7}_{+0.6}$    		&$1.6^{+0.2}_{-0.2}$		&$8.6^{-0.3}_{+0.4}$	\\
\hline
Total             &$100.8^{+3.1}_{-3.0}$   	&$108.8^{-1.1}_{+0.8}$ 		&$63.3^{+0.8}_{-0.5}$		&$152.3^{+1.8}_{-0.8}$	\\
\hline
$\frac{\mathcal{B}(D^*K)}{\mathcal{B}(D^{\unseen{*}}K)}$
			  &$0.91^{-0.18}_{+0.18}$ 		&$0.62^{+0.22}_{-0.12}$ 		&$0.91^{-0.18}_{+0.18}$	&$1.09^{+0.15}_{-0.11}$	\\
\hline
\end{tabular}
\end{table}

If we do not restrict the mass of $2^3\!S_1$ state is less than that of $1^3\!D_1$ state, we can obtain another large mixing angle $\theta_s\simeq -77\degree$, which could also reproduce the experiment ratio $R_K[D^*_{s1}(2710)]=0.91$. The strong decay properties are also listed in \autoref{Tab-Ds-width2}. In such case, the $\Gamma[D^*_{s1}(2700)]$ is $\sims63$ \si{MeV}, which is about the half of experimental value; the total width for its orthogonal partner $D^*_{s1}(2860)$ is about 150 \si{MeV}; the ratio $R_K[D^*_{s1}(2860)]$ is 1.09. Refs.\,\cite{XHZ-2010,DML-2010,DML-2011} also achieved a large mixing angle $-(57\sims77)\degree$. We noticed that in Ref.\,\cite{God-2014} the large mixing angle is also obtained, although in the later work\,\cite{God2-2014} the authors denied this possibility.

\begin{table}[ht]
\setlength{\tabcolsep}{5pt}
\caption{Comparisons with other Refs. when taking $D^*_{s1}(2700)$ and $D^*_{s1}(2860)$ as the mixtures of $2{^3\!S_1}$-$1{^3\!D_1}$ $c\bar s$. Decay width $\Gamma$ is in unit of \si{MeV} and the mixing angle $\theta_s$ is in unit of degree.}\label{Tab-Ds-w2}
\vspace{0.2em}\centering
\begin{tabular}{ |c|c|c|c|c|c|c|c| }
\hline
Mode		          	  	&Exp.			& This   				&Ref.\,\cite{DML-2011} &Ref.\,\cite{QTS-2015}	& Ref.\,\cite{BC-2015}\\
\hline
$\theta_s $ 		     &-			& $8.7^{+3.9}_{-3.2}$   	&-(61\sims77)   	  & 6.8\sims11.2  	&-$(4\sims16)$			\\			\hline 		
$\Gamma_{\scriptstyle D^*_{s1}(2700)}$		
					&$117\pm13$	& $100.8^{+3.1}_{-3.0}$  	&180\sims198   		  & \sims100		&\sims(210\sims220)		\\
$R_K [D^*(2700)]$		&$0.91\pm0.18$	& $0.91^{-0.18}_{+0.18}$ 	&1.16\sims0.66 		  & \sims0.91		&$\sims(1.35\sims 0.69)$	\\
\hline
$\Gamma_{\scriptstyle D^*_{s1}(2860)}$		
					&$159\pm80$	& $108.8^{-1.1}_{+0.8}$  	&40\sims70   		  & \sims300		&\sims(120\sims150)		\\
$R_K [D^*_{s1}(2860)]$	&-			& $0.62^{+0.22}_{-0.12}$ 	&0.04\sims2.71 		  & 0.6\sims0.8		&0.31\sims1.16   		\\			  		
\hline
\end{tabular}
\end{table}

As a short summary, based on our results of the strong decays, we find that, the $2^3\!S_1$-$1^3\!D_1$ mixing scheme with a small mixing angle $\theta_s\simeq8.7\degree$ can well describe the observed the $D^*_{s1}(2700)$ and $D^*_{s1}(2860)$. 
The weak mixing between $2^3\!S_1$ and $1^3\!D_1$ charm-strange mesons is also favored by Refs.\,\cite{BC-2011,God2-2014,QTS-2015,BC-2015}. 

\subsection{$1^-$ charm mesons}
As just stated in the introduction, there are four potential $1^-$ resonances observed in experiments recently, namely, $D^*(2600)$\,\cite{BaBar-2010}, $D^*(2650)$\,\cite{LHCb-2013}, $D^*_1(2680)$\,\cite{LHCb-2016} and $D^*_1(2760)$\,\cite{LHCb-2015}.
The discrepancies among these current experimental data make the classifications more complicated than that for the corresponding charm-strange mesons. LHCb reported two $1^-$ states charm mesons, $D^*_1(2760)$\,\cite{LHCb-2015} and $D^*_1(2680)$\,\cite{LHCb-2016}. Both the two resonances have the same spin-parity $J^P=1^-$. The detected total widths are almost the same, while the mass differences are $\sims100~\si{MeV}$. Besides the two spin-parity determined $1^-$ state $c\bar u$, there is still two natural parity charm mesons $D^*(2600)$\,\cite{BaBar-2010} and $D^*(2650)$\,\cite{LHCb-2013}, whose masses locate in the mass region of $2^3\!S_1$ state $c\bar u$ predicted by the GI model\,\cite{God-1985,God-2016}. However, the measured total widths of $D^*(2600)$ and $D^*(2650)$ are inconsistent by $\sims50$ \si{MeV}.
\begin{table}[ht]
\setlength{\tabcolsep}{3pt}
\caption{Decay properties of $D^*(2600)$, $D^*(2650),~D^*_1(2680)$ and $D^*_1(2760)$ as the $2{^3\!S_1}$ or $1{^3\!D_1}$ dominant $c\bar u$ states in unit of $\si{MeV}$.}\label{Tab-D-width1}
\vspace{0.2em}\centering
\begin{tabular}{|l|c|c|c|c|c|c|c|c|}
\hline
\multirow{2}{*}{Mode}       &\multicolumn{4}{c|}{$2\,^3\!S_1$}		& \multicolumn{4}{c|}{$1^3\!D_1$}  \\                                                           
\cline{2-9}
                           &$D^*(2610)$	&$D^*(2650)$	&$D^*_1(2680)$	&$D^*_1(2760)$	&$D^*(2610)$ &$D^*(2650)$  &$D^*_1(2680)$   &$D^*_1(2760)$ \\

\hline
$D^{*0} \pi^0$              &12.6		&15.3			&17.4			&25.5			&2.9		  &3.6	     &4.2       & 5.8 \\
$D^{0} \pi^0$               &6.7		&7.4			&7.9			&8.6			&13.6		  &15.7		&17.3		&23.2		\\
$D^{*+} \pi^-$              &24.8		&30.3			&34.6			&51.0			&5.5		  &7.0		&8.2		&11.3		\\
$D^{+} \pi^-$               &13.5		&15.0			&16.0			&17.5			&26.6		  &30.8		&34.0		&46.1		\\
$D^{*+}_s K^-$              &0.1		&2.1			&4.8			&20.0			&0		  &0.2		&0.6		&2.2		\\
$D^{+}_s K^-$               &5.4		&7.9			&10.0			&16.6			&5.5		  &8.4		&10.9		&22.1		\\
$D^{*0} \eta$               &1.6		&3.6			&5.2			&11.5			&0.3		  &0.7		&1.0		&2.2		\\
$D^{0} \eta$                &3.7		&4.5			&5.0			&5.9			&5.3		  &6.8		&7.9		&12.0		\\
$D_1^0 \pi^0$               &0.1		&0.4			&0.8			&4.4			&13.0		  &22.6		&30.5		&55.5		\\
$D_1^+\pi^-$                &0.1		&0.6			&1.4			&8.3			&23.6		  &42.9		&58.6		&109.1	\\
$D_1^{\prime 0}\pi^0$       &0.7		&0.6			&0.3			&0.7			&0		  &0.002		&0.005	&0.04		\\
$D_1^{\prime +}\pi^-$       &1.4		&1.2			&0.7			&1.2			&0		  &0.004		&0.008	&0.07		 \\
\hline
Total                       &70.7		&88.9			&104.1		&171.2		&96.3  	  &138.7		&173.2	&289.6	\\
\hline
$\frac{\Gamma(D^+\pi^-)}{\Gamma(D^{*+}\pi^-)}$	
   	 				 &0.54		&0.50			&0.46			&0.34			&4.84		  &4.40		&4.15		&4.08		\\  		
\hline
$\frac{\Gamma(D^+_1\pi^-)}{\Gamma(D^{*+}\pi^-)}$	   	
					 &0.004 		&0.02 		&0.04 		& 0.16		&4.3 		  &6.1 		&3.7 		& 9.7   \\		
\hline
\end{tabular}
\end{table}

Above all, we calculate the strong decays properties by taking all these four resonances as the $2^3\!S_1$ or $1^3\!D_1$ state $c\bar u$. The obtained results are listed in \autoref{Tab-D-width1}. We can see that, when taking $D^*(2600)$ as the $2^3\!S_1$ state, both the total widths and ratio $R_{D^+}[D^*(2600)]$ are comparable with the BaBar measurements, $93$ \si{MeV} and $0.32$\,\cite{BaBar-2010}. Since for the other three resonances, only the total widths can be used to compare with experiments. From the calculated total widths, we can only make rough judgments, both the $2^3\!S_1$ and $1^3\!D_1$ assignments seem reasonable for $D^*(2650)$ and $D^*_1(2680)$, while also considering the mass predictions\,\cite{God-1985,God-2016}, they are more likely to be the $2^3\!S_1$ states. Taking $D^*_1(2760)$ as $1^3\!D_1$ state $c\bar u$ gives the width $\sims 290$ \si{MeV}, which is about $100$ \si{MeV} larger than the LHCb measurement $\sims180$ \si{MeV}\,\cite{LHCb-2015}. We also find that the decay channel $D_1^+\pi^-$ becomes quite important in the decay of $1^3\!D_1$ state, hence we define ratio $R_{D^+_1}=\frac{\Gamma({D^+_1} \pi^-)}{\Gamma(D^{*+}\pi^-)}$. This ratio is quite sensitive to the assignments of $2^3\!S_1$ or $1^3\!D_1$. All in all, for identification of these excited $1^-$ resonances, the consistent measurements from experiments are necessary and pivotal.

$D^*(2600)$ seems consistent with the $2^3\!S_1$ assignment, while the predicted ratio $\frac{\Gamma({D^+} \pi^-)}{\Gamma(D^{*+}\pi^-)}=0.54$ is a little larger than BaBar measurement 0.32\,\cite{BaBar-2010}. This small discrepancy between theoretical and experimental results hints, there exists a small mixing between the $2^3\!S_1$ and $1^3\!D_1$ states. The physical quantity $R_{D^+} [D^*(2600)]$ can behave as a good restriction to the mixing angle, just as what we have done in the $1^-$ charm-strange systems. Then again we introduce the $2^3\!S_1$-$1^3\!D_1$ mixing scheme as
\begin{equation}
\begin{bmatrix}|D^{*}_{a}\rangle\\|D^*_{b} \rangle  \end{bmatrix}= \begin{bmatrix}\cos\theta_u & \sin\theta_u \\ -\sin\theta_u& \cos\theta_u \end{bmatrix} \begin{bmatrix}|2^3\!S_1\rangle \\ |1^3\!D_1\rangle \end{bmatrix}.
\end{equation}

\begin{figure}[ht]
\centering
\subfigure[Total width $\Gamma_{D^*_b}$ versus the mass.]{\includegraphics[width=0.48\textwidth]{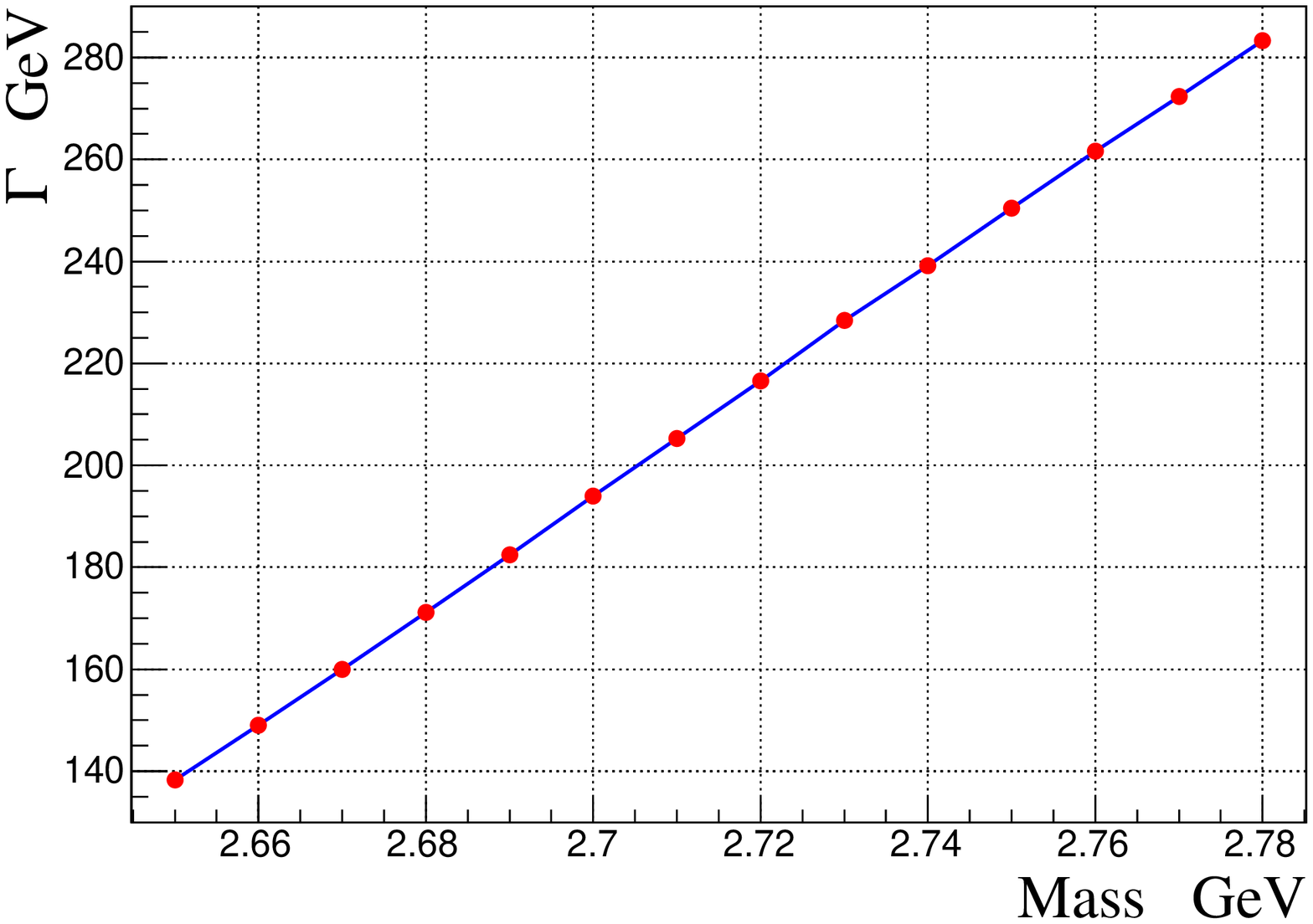} \label{Fig-w-mass}}
\subfigure[$R_{D^+}(D^*_b)$ versus the mass.]{\includegraphics[width=0.48\textwidth]{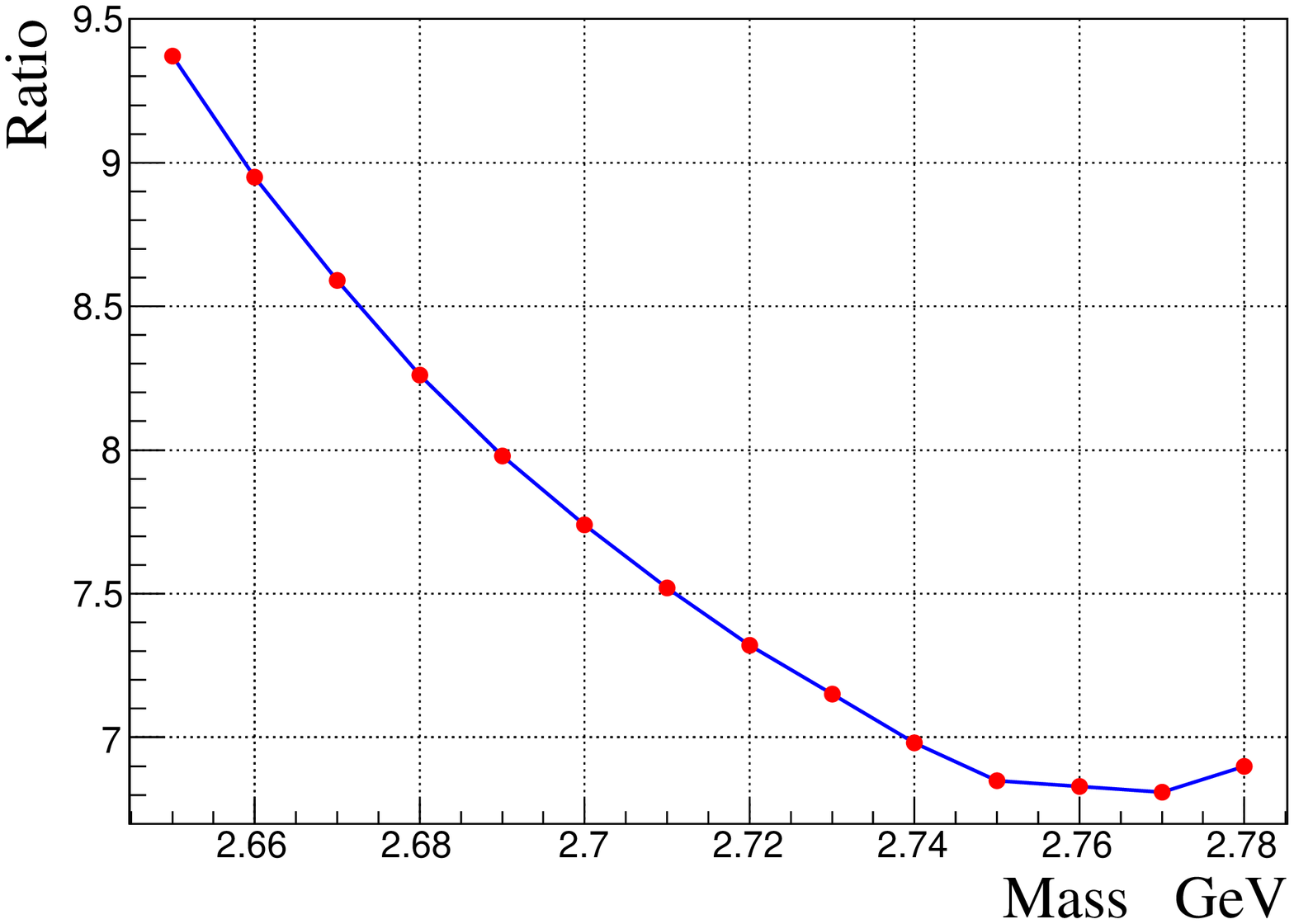} \label{Fig-r-mass}}\\
\caption{The variation of $\Gamma_{D^*_b}$ and ratio $R_{D^+}(D^*_b)$ change along with the mass of $D^*_b$, where $D^*_b$ is the state dominant by $1^3\!D_1$ component and $R_{D^+}(D^*_b)=\frac{\Gamma(D^*_b \to D^+\pi^-)}{\Gamma(D^*_b \to D^{*+}\pi^-)}$.}\label{Fig-rw-mass}
\end{figure}

At first, we take $D^*(2600)$ as the $1^-$ state $c\bar u$ dominant by $2\,^3\!S_1$ components, while $D^*(2650)$ as the orthogonal partner of $D^*(2600)$. To fix the ratio $R_{D^+}[D^*(2600)]$ at BaBar's measurement $0.32^{-0.09}_{+0.09}$\,\cite{BaBar-2010}, we obtain the mixing angle $\theta_u=-(7.5^{+4.0}_{-3.3})\degree$. The theoretical uncertainties are induced by varying the experimental ratio $R_{D^+}{[D^*(2600)]}$ in $1\sigma$ range of its central value. Our results reveal that the mixing angle $\theta_u$ is not sensitive to the mass of $D^*_b$. When $m_{D^*_b}$ ranges from $2.65$ to $2.78$ \si{GeV}, the variation of mixing angle is about $0.1$ degree. So we will ignore this tiny difference in the following statements. The partial decay widths are listed in \autoref{Tab-D-width2}, where $D^*(2650),~D^*(2680)$ or $D^*_1(2760)$ is taken as the orthogonal partner of $D^*(2600)$. The dependence of $\Gamma_{D^*_b}$ and ratio $R_{D^+}(D^*_b)$ over the mass of $D^*_b$ can be seen in \autoref{Fig-rw-mass}, where we let $m_{D^*_b}$ range from $2.65$ to $2.78$ \si{GeV}. It can be seen clearly in \autoref{Fig-w-mass} that the corresponding $\Gamma_{D^*_b}$ increases from  about 140~\si{MeV} to 290 \si{MeV}. The predicted ratio $R_{D^+}{[D^*_b]}$ goes down from 9.4 to 6.8, which is displayed in \autoref{Fig-r-mass}. The calculated total width $\Gamma_{D^*(2600)}=66$ \si{MeV} is comparable with BaBar's measurement $93$ \si{MeV}\,\cite{BaBar-2010}, while $\Gamma_{D^*_b}$ locates in the range $142\sims291$ \si{MeV} when $m_{D^*_b}$ varies from $2.65$ to $2.78$ \si{GeV}.

\begin{table}[ht]
\caption{The strong decay properties with the further $2^3\!S_1$-$1^3\!D_1$ mixing scheme, where $D^*(2600)$ is taken as the $D^*_a$ state and $D^*(2650)$, $D^*_1(2680)$ or $D^*_1(2760)$ is taken as the $D^*_b$ state. The unit for decay width is in \si{MeV}. The obtained mixing angle $\theta_u=-(7.5^{+4.0}_{-3.3})\degree$ when the ratio $R_{D^+}{[D^*(2600)]}$ ranges in $1\sigma$.}\label{Tab-D-width2}
\vspace{0.2em}\centering
\begin{tabular}{|l|r|r|r|r|}
\hline
\multirow{2}{*}{Mode}     					&$m[D^*_a]$		& \multicolumn{3}{c|}{$m[D^{*}_{b}]$}  \\                                                           
\cline{2-5}
                            & 2610				& 2650      			&2680    				&2780 \\
\hline
$D^{*0} \pi^0$              &$14.0^{+0.7}_{-0.6}$	& $2.0^{-0.7}_{+0.7}$       &$2.4^{-0.7}_{+0.8}$     & $3.7^{-0.9}_{+0.9}$ \\
$D^{0} \pi^0$               &$4.4^{-1.1}_{+1.0}$	& $18.1^{+1.1}_{-1.1}$      &$19.7^{+1.0}_{-1.0}$    & $25.3^{+0.8}_{-0.8}$ \\
$D^{*+} \pi^-$              &$27.6^{+0.2}_{-1.2}$	& $3.8^{-1.3}_{+1.4}$       &$4.7^{-1.5}_{+1.5}$     & $7.3^{-1.9}_{+1.7}$ \\
$D^{+} \pi^-$               &$8.8^{-2.2}_{+2.0}$	& $35.6^{+2.2}_{-2.0}$      &$38.8^{+2.2}_{-2.0}$    & $50.4^{+1.7}_{-1.8}$ \\
$D^{*+}_s K^-$              &$0.1^{+0}_{-0}$		& $0.1^{-0.1}_{+0.1}$       &$0.3^{-0.2}_{+0.1}$     & $1.1^{-0.4}_{+0.5}$  \\
$D^{+}_s K^-$               &$4.0^{-0.7}_{+0.6}$	& $10.3^{+1.0}_{-0.8}$      &$13.3^{+1.2}_{-1.0}$    & $25.6^{+1.7}_{-1.4}$ \\
$D^{*0} \eta$               &$1.8^{+0.0}_{-0.0}$	& $0.3^{-0.1}_{+0.0}$		 &$0.5^{-0.1}_{+0.1}$     & $1.2^{-0.2}_{+0.2}$  \\
$D^{0} \eta$                &$2.6^{-0.3}_{+0.2}$	& $8.1^{+0.3}_{-0.3}$       &$9.4^{+0.4}_{-0.3}$     & $13.6^{+0.3}_{-0.4}$\\
$D_1^0 \pi^0$               &$0.4^{+0.4}_{-0.4}$	& $22.1^{-0.6}_{+0.3}$      &$29.7^{-0.7}_{+0.5}$    & $54.8^{-1.1}_{+0.5}$  \\
$D_1^+\pi^-$                &$0.7^{+0.7}_{-0.3}$	& $41.9^{-1.2}_{+0.6}$      &$57.2^{-1.5}_{+0.8}$    & $107.5^{-2.2}_{+1.1}$  \\
$D_1^{\prime 0}\pi^0$       &$0.7^{-0.02}_{+0.01}$	& $0.02^{+0.03}_{-0.01}$    &$0.03^{+0.02}_{-0.02}$  & $0.04^{+0.01}_{-0.0}$\\
$D_1^{\prime +}\pi^-$       &$1.3^{-0.0}_{+0.1}$	& $0.05^{+0.05}_{-0.03}$    &$0.05^{+0.05}_{-0.03}$  & $0.07^{+0.02}_{-0.01}$  \\
\hline
 $\Gamma_\up{Total}$        &$66.4^{-2.3}_{+1.4}$	&$142.4^{+0.7}_{-1.1}$      &$176.1^{+0.2}_{-0.9}$   & $ 290.6^{-2.2}_{+0.5}$\\
\hline
$\frac{\Gamma(D^+\pi^-)}{\Gamma(D^{*+}\pi^-)}$	
					 &$0.32^{-0.09}_{+0.09}$	&$9.4^{+5.8}_{-2.9}$		 &$8.3^{+4.6}_{-2.3}$     & $6.9^{+2.7}_{-1.5}$\\
\hline
\end{tabular}
\end{table}

\begin{table}[ht]
\caption{The strong decay properties with the further $2^3\!S_1$-$1^3\!D_1$ mixing scheme, where $D^*(2650)$ is taken as the $D^*_a$ state and the mass of $D^*_b$ is taken as $2.68$, $2.73$ and $2.78$ \si{GeV}, respectively. We assume that the ratio $\frac{\mathcal{B}(D^{+ }\pi^-)}{\mathcal{B}(D^{*+}\pi^-)}=0.32\pm0.09$ as that for $D^*(2600)$. Our obtained mixing angle $\theta_u=-(6.1^{+4.0}_{-3.4})\degree$. The unit for decay width is in \si{MeV}.}\label{Tab-D-width2b}
\vspace{0.2em}\centering
\begin{tabular}{|l|r|r|r|r|}
\hline
\multirow{2}{*}{Mode}     					&$m[D^*_a]$		& \multicolumn{3}{c|}{$m[D^{*}_{b}]$}  \\                                                           
\cline{2-5}
                            & 2650				& 2680      			 &2730    				&2780 \\
\hline
$D^{*0} \pi^0$              &$16.7^{+0.8}_{-0.7}$  & $2.6^{-0.8}_{+0.9}$   & $3.4^{-1.0}_{+1.0}$   & $3.9^{-1.0}_{+1.0}$ \\
$D^{0} \pi^0$               &$5.2^{-1.3}_{+1.2}$   & $19.5^{+1.2}_{-1.2}$  & $22.2^{+1.2}_{-1.2}$  & $25.3^{+1.1}_{-1.1}$ \\
$D^{*+} \pi^-$              &$33^{+1.6}_{-1.4}$    & $5.1^{-1.7}_{+1.7}$   & $6.6^{-2.0}_{+1.9}$   & $7.7^{-2.1}_{+1.9}$ \\
$D^{+} \pi^-$               &$10.6^{-2.7}_{+2.3}$  & $38.4^{+2.6}_{-2.3}$  & $44^{+2.5}_{-2.3}$    & $50.3^{+2.3}_{-2.2} $\\
$D^{*+}_s K^-$              &$2.2^{+0.03}_{-0.09}$ & $0.3^{-0.2}_{+0.1}$   & $0.7^{-0.3}_{+0.4}$   & $1.2^{-0.5}_{+0.5} $\\
$D^{+}_s K^-$               &$6.2^{-1.1}_{+1.0}$   & $13^{+1.4}_{-1.1}$    & $18.6^{+1.6}_{-1.5}$  & $25.4^{+1.9}_{-1.8} $\\
$D^{*0} \eta$               &$3.9^{+0.1}_{-0.2}$   & $0.6^{-0.2}_{+0.2}$   & $1^{-0.3}_{-0.4}$     & $1.4^{-0.5}_{+0.4} $\\
$D^{0} \eta$                &$3.3^{-0.7}_{+0.7}$   & $9.2^{+0.8}_{-0.7}$   & $11.3^{+0.8}_{-0.8}$  & $13.5^{+0.8}_{-0.8}$ \\
$D_1^0 \pi^0$               &$0.8^{+0.5}_{-0.3}$   & $29.9^{-0.7}_{+0.4}$  & $43^{-0.9}_{+0.4}$    & $55^{-1.0}_{+0.4} $\\
$D_1^+\pi^-$                &$1.4^{+1.0}_{-0.5}$   & $57.6^{-1.4}_{+0.7}$  & $83.7^{-1.8}_{+0.9}$  & $107.9^{-2.0}_{+0.9} $\\
$D_1^{\prime 0}\pi^0$       &$0.6^{-0.02}_{+0.0}$  & $0.01^{+0.02}_{-0.0}$ & $0.02^{+0.01}_{-0.0}$ & $0.04^{-0.04}_{-0.0} $\\
$D_1^{\prime +}\pi^-$       &$1.2^{+0.0}_{-0.0}$   & $0.03^{+0.03}_{-0.02}$& $0.04^{+0.02}_{-0.01}$& $0.06^{+0.05}_{+0.01}$ \\
\hline
 $\Gamma_\up{Total}$        &$85.1^{-1.7}_{+1.9}$  & $176.2^{+1.0}_{-1.4}$ & $234.6^{-0.2}_{-0.9}$ & $291.7^{-0.9}_{-0.8}$ \\
\hline
$\frac{\Gamma(D^+\pi^-)}{\Gamma(D^{*+}\pi^-)}$	
					 &$0.32^{-0.09}_{+0.09}$ & $7.53^{+4.56}_{-2.19}$  & $6.67^{+3.28}_{-1.76}$  & $6.53^{+2.89}_{-1.52}$  \\
\hline
\end{tabular}
\end{table}
Certainly, several other mixing schemes and corresponding assignments are still possible, however, there is no experimental ratio like $R_{D^+}{[D^*(2600)]}$ for $D^*(2650)$, $D^*_1(2680)$ or $D^*_1(2760)$ to restrict the mixing angle. In \autoref{Tab-D-width2b}, the decay properties are displayed when taking $D^*(2650)$ as the $|D^*_a\rangle$ state with varying $m_{D^*_b}$ from $2.68$ to $2.78$ \si{GeV}, where we still assume the ratio $R_{D^+}[D^*(2650)]=0.32\pm0.09$ for $D^*(2650)$. In this case, we find the mixing angle $\theta_{u}=-(6.1^{+4.0}_{-3.4})\degree$, $\Gamma_{D^*(2650)}=85.1$ \si{MeV} and $\Gamma_{D^*_b}$ ranges in $176\sims292$ \si{MeV} when $m_{D^*_b}$ varies from $2.68$ to $2.78$ \si{GeV}. Of course, we can still take $D^*_1(2680)$ as the $|D^*_a\rangle$ state while $D^*_1(2760)$ as the $|D^*_b\rangle$ state, the results should behave similar with above tests, the mixing angle will be even smaller, and the corresponding properties will behave almost the same with that under the assignments without this further mixing.

The comparisons of our results with others can be found in \autoref{Tab-D-width3}, where we have also listed the properties of $D^*_1(2760)$ when taken it as the $|D^*_b\rangle$ state in order to make a comparison. From \autoref{Tab-D-width3}, we can see that both our small mixing angle and $\Gamma_{D^*(2600)}$ are consistent with other predictions, except for total width in Ref.\,\cite{BC-2015}, which is about 3 times larger than ours. Also should be noticed that, the ratio $R_{D^+}(D^*_b)$ is sensitive to the variation of mixing angle $\theta_u$.

Based on our calculations and current experimental results, it is still difficult to make definite assignments to the observed $D^*(2600)$, $D^*(2650)$, $D^*_1(2680)$ or $D^*_1(2760)$. For these excited $1^-$ charm states, $D^{(*)}\pi$ channels are the important decay modes for both $2^3\!S_1$ and $1^3\!D_1$ assignments, and can amount to $(60\sims80)\%$ and $(30\sims50)\%$ of the total widths, respectively. Besides, we can still make the following summary:
\begin{enumerate}
\item Decay channels $D_1\pi$ become very important for $1^3\!D_1$ state $c\bar u$ and can amount to about $50\%$ among the total width, while these decay modes can be ignored in the $2^3\!S_1$ state. This feature can be used to determined the essence of these $1^-$ charm resonances.
\item The properties of $D^*(2600)$ reveal it is predominant by the $2^3\!S_1$ component. The BaBar measured ratio $\frac{\mathcal{B}(D^{+ }\pi^-)}{\mathcal{B}(D^{*+}\pi^-)}$\,\cite{BaBar-2010} can be explained by a small $2^3\!S_1$-$1^3\!D_1$ mixing. And our obtained mixing is about $-7.5\degree$. This result is consistent with that for the $1^-$ states $D^*_{s1}(2700)$ and $D^*_{s1}(2860)$, where we got a mixing angle $\theta_s=8.7\degree$.
\item The mass of $D^*_1(2760)$ is consistent with the prediction of $1^3\!D_1$ state $c\bar u$\,\cite{God-1985,God-2016}. If we take this assignment, the measured total width seems too small (the LHCb result\,\cite{LHCb-2015} is about $100$ \si{MeV} smaller than theoretical calculation). This conclusion is also favored by the researches in Refs.\,\cite{Zhong-2010,DML-2011,BC-2015,QTS2-2015,God-2016}.
\item $D^*(2650)$ is more likely to be the $2^3\!S_1$ predominant state. There is no great conflict in the total widths of $2^3\!S_1$ and $1^3\!D_1$ assignments, while its mass is more consistent with $2^3\!S_1$ state. The ratio $\frac{\Gamma(D^{+ }\pi^-)}{\Gamma(D^{*+}\pi^-)}$ behaves quite differently for the two assignments, which is $0.5$ for $2^3\!S_1$ assignment and $4.4$ for $1^3\!D_1$ assignment. Hence this ratio can be used to test the essence of $D^*(2650)$.
\item For  $D^*_1(2680)$, the situation is similar with $D^*(2650)$. There exists no great conflict in the total widths between $2^3\!S_1$ and $1^3\!D_1$ assignments compared with experimental measurements. The ratios $\frac{\Gamma(D^{+ }\pi^-)}{\Gamma(D^{*+}\pi^-)}$ are $0.46$ and $4.15$ for $2^3\!S_1$ and $1^3\!D_1$ assignment, respectively, therefore can be used to discriminate the essence of $D^*_1(2680)$.
\end{enumerate}
\begin{table}[H]
\setlength{\tabcolsep}{5pt}
\caption{Comparison with other Refs. when taking $D^*(2600)$ as the mixture of $2{^3\!S_1}$-$1{^3\!D_1}$ $c\bar u$. $\Gamma_\up{Tot}$ is in unit of \si{MeV} and the mixing angle $\theta_u$ is in unit of degree.}\label{Tab-D-width3}
\vspace{0.2em}\centering
\begin{tabular}{ |c|c|c|c|c|c|c| }
\hline
$D^*(2600)$          	  	& Exp.			& This   				&Ref.\,\cite{DML-2011}	& Ref.\,\cite{Zhong-2010}  &Ref.\,\cite{BC-2015}	& Ref.\,\cite{QTS2-2015}\\
\hline
$\theta_u $ 		     & -				& -$(7.5^{+4.0}_{-3.3})$   &-(21\sims23)   	     &-$(36\pm 6)$			&(4\sims17)			&(-3.6\sims1.8) \\			  		
$\Gamma_{\scriptstyle D^*(2600)}$		
					&$93\pm14.3$ 		& $66.4^{-2.3}_{+1.4}$  	&74\sims80   			&75\sims115			&205\sims195			&$\sims 60$	 \\
$R_{D^+}{[D^*(2600)]}$	
			  		&$0.32\pm0.09$		& $0.32^{-0.09}_{+0.09}$ 	&0.38\sims0.43 			&$0.63\pm 0.21$			&\sims(0.25\sims0.53)		&$\sims 0.32$\\
$\Gamma_{\scriptstyle D^*_1(2760)}$		
					&$177\pm38.4$		& $ 290.6^{-2.2}_{+0.5}$  	&280\sims310   			&300\sims550			&\sims290				&$ 385$	 \\
$R_{D^+}{[D^*_1(2760)]}$	
			  		&-				& $6.9^{+2.7}_{-1.5}$ 	&1.25\sims2.25 			&-   					&(2.62\sims28.86)		&$2.2$\\			  		
\hline
\end{tabular}
\end{table}

\section{Summary}\label{Sec-4}
In this work, we carried out a systematical research on the potential $1^-$ open charm mesons, including the charm-strange mesons $D^*_{s1}(2700)$ and $D^*_{s1}(2860)$, charm mesons $D^*(2600)$, $D^*(2650)$, $D^*_1(2680)$ and $D^*_1(2760)$. The main strong decay properties by taking these natural spin-parity resonances as the $2^3\!S_1$ or $1^3\!D_1$ states are achieved by using the Bethe-Salpeter methods. In particularly, the further $2S$-$1D$ mixing scheme is used to explain both the $1^-$ charm and charm-strange mesons. The obtained results and predicted properties can be tested in the near future experiments.

Our results reveal that, $D^*_{s1}(2700)$ and $D^*_{s1}(2860)$ can be well described by the further $2^3\!S_1$-$1^3\!D_1$ mixing scheme with a small mixing angle $(8.7^{+3.9}_{-3.2})\degree$. Both the total widths and ratio of corresponding partial decay widths are consistent with the experimental measurements. Our predicted ratio $\frac{\Gamma(D^*K)}{\Gamma(D^{\color{white}*}K)}$ for $D^*_{s1}(2860)$ is $0.62^{+0.22}_{-0.12}$, which could be used to test this $2S$-$1D$ mixing scheme in future.
For the corresponding charm mesons, since the experimental measurements are not consistent with each other, the identification and assignments are much more difficult. Based on our results, the BaBar $D^*(2600)$\,\cite{BaBar-2010} can be explained by the same mixing scheme with a mixing angle of $-(7.5^{+4.0}_{-3.3})\degree$. $D^*(2650)$\,\cite{LHCb-2013} and $D^*_1(2680)$\,\cite{LHCb-2016} are more likely to be $2^3\!S_1$ predominant states, since their masses are consistent with the $2^3\!S_1$ predictions, while our calculated total widths are both comparable with the experimental measurements under the $2^3\!S_1$ or $1^3\!D_1$ assignments.  
Our results also show that, the measured width of $D^*_1(2760)$ is much smaller than the theoretical calculations under the $1^3\!D_1$ assignment. This would be an obstacle to identify $D^*_1(2760)$ as the $1^3\!D_1$ predominant $c\bar u$. There still exit puzzles and difficulties in identifications of these new excited charmed mesons. Further precise measurements of their properties are needed and important.

\section*{Acknowledgments}
This work was supported in part by the National Natural Science
Foundation of China (NSFC) under Grant Nos.~11405037, 11575048, 11505039, 11447601, 11535002 and 11675239, and also in part by PIRS of HIT Nos.~T201405, A201409, and B201506.

\end{document}